\documentclass[useAMS,usenatbib]{mn2e}
\usepackage{amsmath}
\usepackage{graphicx}
\title[SMBH formation by cold accretion shocks]
{Supermassive black hole formation 
by the cold accretion shocks in the first galaxies}
\author[K. Inayoshi and K. Omukai]{Kohei Inayoshi$^{1}$
\thanks{E-mail:inayoshi@tap.scphys.kyoto-u.ac.jp} 
and Kazuyuki Omukai$^{1}$
\thanks{E-mail:omukai@tap.scphys.kyoto-u.ac.jp}
\\
$^{1}$Department of Physics, Graduate School of Science, Kyoto University, 
Kyoto 606-8502, Japan }
\begin{document}

\date{}

\pagerange{000--000} \pubyear{0000}

\maketitle

\label{firstpage}

\begin{abstract}
We propose a new scenario for supermassive star (SMS; $\ga 10^5$ M$_{\sun}$) 
formation in shocked regions of colliding cold accretion flows 
near the centers of first galaxies.
Recent numerical simulations indicate that 
assembly of a typical first galaxy with virial temperature 
$T_{\rm vir} \ga 10^{4}$ K proceeds via cold and dense flows penetrating 
deep to the center, where the supersonic streams 
collide each other to develop a hot ($\sim 10^4$ K) 
and dense ($\sim 10^3$ cm$^{-3}$) shocked gas.
The post-shock layer first cools by efficient Ly$\alpha$ emission
and contracts isobarically until $\simeq 8000$ K.
Whether the layer continues the isobaric contraction depends 
on the density at this moment:    
if the density is high enough for collisionally 
exciting H$_2$ rovibrational levels ($\ga 10^{4}$ cm$^{-3}$), 
enhanced H$_2$ collisional dissociation 
suppresses the gas to cool further.
In this case, the layer fragments into massive 
($\ga 10^5$ M$_{\sun}$) clouds, which collapse 
isothermally ($\sim 8000$ K) by the Ly$\alpha$ cooling 
without subsequent fragmentation.
As an outcome, SMSs are expected to form and evolve eventually 
to seeds of supermassive black holes (SMBH).
By calculating thermal evolution of the post-shock 
gas, we delimit the range of post-shock conditions 
for the SMS formation, which can be expressed as: 
$T\ga 6000$ K $(n_{\rm H}/10^4~{\rm cm}^{-3})^{-1}$
for $n_{\rm H} \la 10^4~{\rm cm}^{-3}$ and 
$T\ga 5000-6000$ K for $n_{\rm H} \ga 10^4~{\rm cm}^{-3}$,
depending somewhat on initial ionization degree.
We found that metal enrichment does not affect the above condition
for metallicity below $\simeq 10^{-3}$ Z$_{\sun}$ if metals are in the gas phase,
while condensation of several percent 
of metals into dust decreases this critical value 
of metallicity by an order of magnitude.
Unlike the previously proposed scenario for SMS formation, 
which postulates extremely strong ultraviolet radiation 
to quench H$_2$ cooling, our scenario here 
naturally explains the SMBH seed formation in the assembly process 
of the first galaxies, even without such a strong radiation. 
\end{abstract}


\section{Introduction}
Discovery of high-$z$ quasars has demonstrated 
the existence of supermassive black holes 
(SMBH) with $\sim 10^9$ M$_{\sun}$ already at the age of the universe 
$\la 1$ Gyr (e.g., Fan 2006; Willott et al. 2007; 
Mortlock et al. 2011).
As an origin of such SMBHs, seed BH formation as a remnant of 
population III stars ($M_{\rm seed} \sim 100$ M$_{\sun}$) and 
their subsequent growth by 
merger and gas accretion have been studied by a number of authors
(e.g., Haiman \& Loeb 2001; Volonteri, Haardt \& Madau 2003; Li et al. 2007).
Even with the Eddington accretion rate $\dot{M}_{\rm Edd}=L_{\rm Edd}/\epsilon c^2$, where $L_{\rm Edd}$ is the Eddington luminosity and $\epsilon \simeq 0.1$ is 
the radiative efficiency, the growth time  
$t_{\rm grow}=0.05\ln (M_{\rm BH}/M_{\rm seed})$ Gyr becomes as long as 
the age of the universe at $z\simeq 6$ ($\simeq 0.8$ Gyr):  
the seed BHs are thus required to keep growing at the Eddington rate.
However, negative feedbacks by the growing BHs prevent such efficient accretion 
(Milosavljevi{\'c}, Couch \& Bromm 2009; Alvarez, Wise \& Abel 2009; Jeon et al. 2011).

As a solution to this, alternative possibility of 
massive seed BH formation by direct collapse of 
supermassive stars (SMS; $\ga 10^5$ M$_{\sun}$) has been considered by some authors. 
Specifically, SMS formation in massive halos 
($T_{\rm vir}\ga 10^4$ K) irradiated 
with strong far ultraviolet (FUV) radiation has often been studied 
(e.g., Bromm \& Loeb 2003; Regan \& Haehnelt 2009a, b; 
Shang, Bryan, \& Haiman 2010).
Since H$_2$ molecule, the main coolant in the primordial gas, 
is photodissociated with strong FUV radiation in the Lyman and Werner bands, 
clouds under such an environment collapse isothermally at $\sim 8000$ K 
by Ly$\alpha $ cooling without fragmentation, if they are massive 
enough with $\ga 10^5$ M$_{\sun}$.
As an outcome of such collapse, SMSs are expected to form.
A massive seed BH as a remnant of SMS collapse reduces 
the growth time to $10^{9}$ M$_{\sun}$ within $0.46$ Gyr
and mitigates the growth-time problem by a big margin.
This scenario, however, has a serious drawback: 
for this mechanism of SMS formation to work, extremely 
strong FUV radiation
$J_{21}^{\rm LW}\ga 10^2-10^3$ (in a unit of $10^{-21}$ erg 
s$^{-1}$ cm$^{-2}$ Hz$^{-1}$ sr$^{-1}$) is required
(Omukai 2001; Bromm \& Loeb 2003; Shang, Bryan \& Haiman 2010), while
the fraction of halos irradiated with such intense FUV fields
with $J_{21}^{\rm LW}\ga 10^3$ is estimated to be $\la 10^{-6}$ at $z\sim 10$ 
(Dijkstra et al. 2008),
i.e., only extremely rare halos satisfy the condition for SMS formation.
Moreover, if high energy components, such as cosmic rays or X-rays, 
are present along with the FUV radiation, their ionization effect promotes 
the H$_2$ formation and then strongly suppresses the SMS formation (Inayoshi \& Omukai 2011).
Although the above scenario might be still viable considering 
the rarity of high-z SMBHs, it is worthwhile to explore another possibility.

In this paper, we propose a new scenario for SMS formation 
in post-shock gas of cold accretion flows in forming first galaxies.
Recent numerical simulations of galaxy formation 
have revealed that, in halos with virial temperature 
$T_{\rm vir}\ga 10^4$ K, the shock position does not stay 
at the virial radius and shrinks inside owing to the efficient 
Ly$\alpha$ cooling, and the accreting cold gas penetrates deep 
to the center through dense filamentary flows 
(Birnboim \& Dekel 2003; Kere{\v s} et al. 2005; Dekel \& Birnboim 2006; 
Dekel et al. 2009; Bromm \& Yoshida 2011). 
The supersonic flows collide each other and the resultant shock 
develops a hot and dense ($\sim 10^4$ K and 
$\sim 10^3$ cm$^{-3}$) gas near the center 
(Wise \& Abel 2007; Greif et al. 2008; Wise, Turk \& Abel 2008).
By studying thermal evolution of the shocked gas, 
we have found that, if the post-shock density is high enough for 
the H$_2$ rovibrational levels to reach the local thermodynamic equilibrium (LTE),  
the efficient collisional dissociation suppresses 
H$_2$ cooling, and the gas cannot cool below several thousand K.
Massive clouds with $\ga 10^5$ M$_{\sun}$ formed by fragmentation of 
the post-shock layer subsequently collapse isothermally at $\sim 8000$ K
by the Ly$\alpha $ cooling. 
Without further fragmentation, monolithic collapse of the clouds 
results in the SMS formation.
Note that, unlike the previous SMS formation mechanism, 
strong FUV radiation is not required in our scenario. 
Similar analysis has also been carried out by 
Safranek-Shrader, Bromm \& Milosavljevi{\'c} (2010), who studied the 
fragmentation of the cold-stream shocked layer 
considering the effects of radiation field and chemical enrichment, but 
for a single post-shock condition 
($4 \times 10^3 {\rm cm^{-3}}$, $1.1 \times 10^4$ K, and the equilibrium 
chemical abundances).

The organization of this paper is as follows.
In Section 2, we describe the model for calculation of thermal evolution 
in the shocked gas.  
In Section 3, we present our results and clarify the conditions for 
isothermal collapse leading to the SMS formation
in terms of post-shock density and temperature. 
Effects of metal enrichment is also considered here.
In Section 4, we analyze thermal processes in the shocked gas 
and discuss the reason for the bifurcation of thermal evolution in more detail. 
Finally, we summarize our study and present some discussions 
in Section 5. 

\section{Model}
In this section, we describe our model for calculation 
of thermal evolution in hot and dense shocked regions 
formed by collision of cold accretion flows in the first galaxies. 

\subsection{Evolution in the post-shock layer}
We consider the thermal evolution in the post-shock layer
under the assumption that the flow is steady and plane-parallel.
Since the post-shock temperature is as high as 
the virial temperature of first-galaxy forming halos ($T_{\rm vir}\ga 10^4$ K), 
cooling by the Ly$\alpha $ emission is efficient early on.
The post-shock flow is compressed almost isobarically 
as long as the gas cools effectively (e.g., Shapiro \& Kang 1987).
Within the steady-state approximation, the conservation of mass 
and momentum leads to the following relationships 
between the density $\rho _0$, pressure $p_0$, 
and flow velocity $v_0$
just behind the shock with those in the post-shock flow
$\rho $, $p$, and $v$:
\begin{equation}
\rho v=\rho _0v_0,
\label{eq:shock1}
\end{equation}
\begin{equation}
\rho v^2+p=\rho _0v_0^2+p_0.
\label{eq:shock2}
\end{equation}
Along this flow, we solve the energy equation
\begin{equation}
\frac{dE}{dt}=\frac{p+E}{\rho }\frac{d\rho }{dt}-\Lambda _{\rm net},
\label{eq:shock3}
\end{equation}
where $E$ is the internal energy per unit volume, 
$d/dt$ is the Lagrangian time derivative, and 
$\Lambda _{\rm net}$ is the net cooling rate per unit volume.
Assuming a strong shock and neglecting the thermal pressure 
in the pre-shock flow, $p_0\simeq 3\rho _0v_0^2$ is satisfied 
just behind the shock front. 
Thus, we approximate the right-hand side of equation (\ref{eq:shock2}) by $4\rho_0v_0^2$.
The cooling term $\Lambda _{\rm net}$ includes
the radiative cooling by H, H$_2$ and HD,  and 
cooling/heating associated with chemical reactions.
We solve the chemical reactions of primordial gas among 
the following 14 species; H, H$_2$, e$^-$, H$^+$, H$^{+}_2$, 
H$^-$, He, He$^+$, He$^{++}$, D, HD, D$^+$, HD$^+$, and D$^-$.
We adopt the same coefficients for the cooling/heating and 
the chemical reactions as in Inayoshi \& Omukai (2011) 
except for omitting the radiative/cosmic-ray ionization 
and dissociation in this calculation.
In studying effects of metal enrichment, 
we add the cooling by the fine-structure-line emission of 
C$_{\rm II}$ and O$_{\rm I}$ to the primordial processes 
described above. 
Assuming the fraction of metals depleted to dust grains 
to be the same as in the Galactic interstellar gas, 
we set the number fractions of 
C and O nuclei in the gas phase with respect to H nuclei to
$x_{\rm C, gas}=0.927\times 10^{-4}(Z /{\rm Z}_{\sun})$ and	
$x_{\rm O, gas}=3.568\times 10^{-4}(Z /{\rm Z}_{\sun})$ 
(Pollack et al. 1994).
We follow Hollenbach \& McKee (1989) in calculating 
the cooling rates of C$_{\rm II}$ and O$_{\rm I}$.
We curtail the C and O chemistry by
simply assuming that all the C and O are in the states 
of C$_{\rm II}$ and O$_{\rm I}$, respectively, from 
the following consideration:
with lower ionization energy ($11.26$ eV) than H atom, 
C is photoionized by weak background radiation and 
in the state of C$_{\rm II}$, 
while O is in ionization equilibrium with H 
and almost neutral for $\la 8000$ K, 
where the O$_{\rm I}$ cooling is important. 
Molecular cooling of metals (e.g., CO and H$_2$O) is not 
included since its cooling is not important 
in the temperature range relevant for the bifurcation of thermal 
evolution ($\ga$ several 10$^3$ K).

Next, we consider the condition for the gravitational instability 
during the isobaric compression of the post-shock layer and thus
for its fragmentation.
For the isobaric compression, the dynamical time
$t_{\rm dyn}(\equiv \rho /(d\rho /dt))$, 
which characterizes thermal evolution, 
is approximately equal to the cooling time 
\begin{equation}
t_{\rm cool}=\frac{(3/2)n_{\rm H}k_{\rm B}T}{\Lambda _{\rm net}},
\end{equation}
where $n_{\rm H}$ is the number density of H nuclei 
and $T$ is the temperature.
On the other hand, the growth timescale for the gravitational 
instability is given by the free-fall time (e.g., Larson 1985)
\begin{equation}
t_{\rm ff}=\sqrt{\frac{32}{3\pi G\rho }}.
\end{equation}

As long as the cooling is effective enough and so 
$t_{\rm cool}(\simeq t_{\rm dyn})\la t_{\rm ff}$,
the post-shock layer continues to be compressed isobarically. 
However, once the cooling becomes ineffective and 
$t_{\rm cool}$ exceeds $t_{\rm ff}$, the contraction of the layer 
halts and a dense layer begins to develop inside the post-shock region.    
For example, since the growth rate of the baryonic mass
in halos of $\sim 10^8$ M$_{\sun}$ at $z\sim 10$ 
is $\sim 4\times 10^{-2}$ M$_{\sun}/{\rm yr}$ (Dekel et al. 2009), 
a gas of $\ga 10^5$ M$_{\sun}$ can accumulate in $\ga 3\times 10^6$ yrs.
If sufficient gas supply is available through the accretion flow, 
the dense layer eventually satisfies the Jeans criterion, i.e., 
the sound-crossing time $t_{\rm cross}$ becomes longer than 
the free-fall time ($t_{\rm cross}\ga t_{\rm ff}$).
The layer then fragments by the gravitational instability
to produce Jeans-scale clouds.
In this paper, assuming that the continuous gas supply to induce the Jeans instability 
is available, we estimate the fragmentation mass scale by the condition 
$t_{\rm ff} \sim t_{\rm cool}$ in accordance with Yamada \& Nishi (1998).

\subsection{Evolution after fragmentation}
After fragmentation, the cloud collapses with its self-gravity, 
and its evolution cannot be modelled as a steady flow anymore.
Density evolution in a cloud collapsing by the self-gravity 
is described by the Penston-Larson self-similar solution 
(Penston 1969; Larson 1969), 
which has density profile with a flat core of 
the Jeans scale and an envelope with the power-law
density distribution $\rho (r)\propto r^{-2}$.
The density in the central core roughly increases  
in the free-fall timescale.
We here calculate the evolution in the central-core part
by using a one-zone model, where the density evolution is
given by 
\begin{equation}
\frac{d\rho }{dt}=\frac{\rho }{t_{\rm ff}}.
\label{eq:density}
\end{equation}
Namely, after the condition for fragmentation 
($t_{\rm cool} \ga t_{\rm ff}$) is satisfied, 
we switch the density evolution described by equations 
(\ref{eq:shock1})-(\ref{eq:shock2}) to that by equation 
(\ref{eq:density}) in our calculation and solve equation (\ref{eq:shock3}) for this density evolution.

When the collapse proceeds significantly and the cloud becomes optically thick, 
the radiative cooling becomes ineffective due to photon trapping. 
We assume the radius of the core to be half a Jeans length
\begin{equation}
R_{\rm c}=\frac{\lambda _{\rm J}}{2}=
\sqrt{\frac{\pi k_{\rm B}T}{G\rho \mu m_{\rm H}}},
\end{equation}
where $\mu $ is the mean molecular weight.
Since we consider the core of the collapsing cloud, 
the column density of i-th species is given by
$N_{\rm i}=x_{\rm i}n_{\rm H}R_{\rm c}$, where $x_{\rm i}$
is its concentration.
Using this value, we estimate the optical depth 
and the reduction rate of radiative cooling as 
in Inayoshi \& Omukai (2011).

\subsection{Initial conditions}
According to numerical simulations of the first galaxy formation
(e.g., Greif et al. 2008; Wise et al. 2008), 
the pre-shock number density and temperature of cold flows
and the shock velocity are typically 
$10^3$ cm$^{-3}$, 200 K, and 20 km s$^{-1}$, 
respectively, 
which correspond to the post-shock density 
$4 \times 10^3$ cm$^{-3}$ and temperature $9000$ K.
With those fiducial values in mind, 
we carry out calculations for 
a wide range of initial number density and temperature: 
$10^2~{\rm cm^{-3}}<n_{\rm H,0}< 10^7~{\rm cm^{-3}}$ and 
$3000~{\rm K}<T_0<10^5~{\rm K}$.
Since H$_2$, the main coolant below $8000$ K, 
forms through the electron-catalyzed reactions
\begin{align}
{\rm H}+{\rm e}^- &\rightarrow {\rm H}^-+\gamma ,\nonumber \\
{\rm H}^-+{\rm H} &\rightarrow {\rm H}_2+{\rm e}^-,
\label{eq:H-}
\end{align} 
the initial ionization degree $x_{\rm e,0}$, along with the initial 
H$_2$ concentration $x_{\rm H_2,0}$, is important quantities 
for the subsequent thermal evolution. 
In reference to the results of Kang \& Shapiro (1992), 
who studied the chemical abundances
in the pre-shock gas considering photo-ionization and dissociation 
by UV radiation emitted from the shock, 
we regard $x_{\rm e,0} \sim 10^{-2}$ and $x_{\rm H_2,0} 
\sim 10^{-6}$ as typical ionization degree and 
molecular fraction, respectively.
However, since cold accretion flows are far denser ($\sim 10^3$ cm$^{-3}$) than the range Kang \& Shapiro (1992) assumed ($\la 10^{-2}$ cm$^{-3}$), the electron recombination 
as well as the shielding of the UV photo-ionization/dissociation 
probably lower the pre-shock ionization degree $x_{\rm e,0}$ 
and elevate molecular fraction $x_{\rm H_2}$ from those values.  
Taking this uncertainty into account, we study the cases 
with a wide range of 
initial ionization degree and H$_2$ fraction:
$10^{-5}\leq x_{\rm e,0}\leq 10^{-1}$ and $10^{-6}\leq x_{\rm H_2,0}\leq 10^{-3}$.

\section{Results}
In this section, we present our results for 
thermal evolution after the gas experiences
the cold accretion shock. 
We first consider the cases of primordial gas and then  
discuss effects of small metal enrichment.

\subsection{Primordial-gas case}
\begin{figure}
\begin{center}
\rotatebox{0}{\includegraphics[height=59mm,width=80mm]{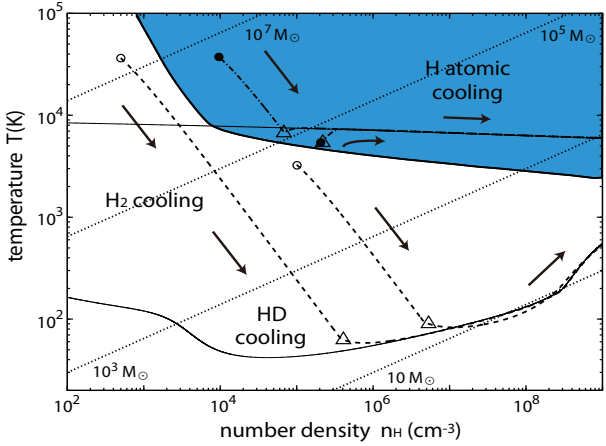}}
\end{center}
\caption{The temperature evolution of primordial gas after heated by 
the cold accretion shock for the initial ionization degree $x_{\rm e,0}=10^{-2}$ 
and H$_2$ fraction $x_{\rm H_2,0}=10^{-6}$.
The evolutionary tracks are shown with dashed and dash-dotted lines 
for four combinations of initial temperature and density, 
indicated by two each open and filled circles. 
From low-density or low-temperature initial conditions 
(dashed lines starting from the open circles), the temperature decreases below 
$\sim 100$ K owing to the H$_2$ and HD cooling.
On the other hand, from dense and hot initial conditions (dash-dotted lines from 
the filled circles), the clouds do not cool below $\sim 8000$ K and 
subsequently collapse almost isothermally by the H atomic cooling.
The triangle symbol on each track indicates 
the epoch when the post-shock layer fragments by the gravitational instability.
The thick solid line, the domain above which is hatched, 
divides the initial conditions leading to 
these two different ways of thermal evolution.
The two thin solid lines show the temperature evolution 
by the H atomic cooling (upper) and the H$_2$ and HD cooling (lower), respectively.
These two evolutionary tracks are calculated by the density evolution of equation (\ref{eq:density})
from the initial conditions $n_{\rm H,0}=10$ cm$^{-3}$, $T_0=10^4$ K, 
and $x_{\rm e,0}=2\times 10^{-4}$ (upper) and $10^{-1}$ (lower), respectively.
For the upper track, the H$_2$ cooling rate is set to zero.
The diagonal dotted lines (lower-left to upper-right) indicate 
the constant Jeans masses, whose values are denoted by numbers in the Figure.
}
\label{fig:region}
\end{figure}

In Fig.~\ref{fig:region}, we show the temperature evolution of 
primordial gas for four post-shock conditions, indicated by   
two each open and filled circles.
We here set the initial ionization degree and 
molecular fraction to $x_{\rm e,0}=10^{-2}$ and $x_{\rm H_2,0}=10^{-6}$, respectively.
First, we see the cases from the open-circle initial conditions in Fig. \ref{fig:region}, whose 
temperature evolution is shown by the dashed lines.
In the lower initial-density case
($n_{\rm H, 0}$, $T_0$)=($5\times 10^2$ cm$^{-3}$, $3.6\times 10^4$ K) among them, 
the post-shock gas cools by the Ly$\alpha $ emission and is compressed isobarically.
Although the Ly$\alpha $ cooling becomes inefficient below $8000$ K, 
enough H$_2$ for cooling has already been formed by this time, which 
enables further temperature decrease. 
HD is formed abundantly for $\la 150$ K, whose cooling eventually 
lowers the temperature to $\sim 50$ K. 
At this point, without efficient coolant anymore, $t_{\rm cool}$
becomes longer than $t_{\rm ff}$.  
Clouds with Jeans mass of several $10$ M$_{\sun}$ are produced 
by the gravitational instability.
Also, in the case of lower initial temperature ($10^5$ cm$^{-3}$, 
 $3.3\times 10^3$ K), abundant H$_2$ is formed immediately.
The cooling by H$_2$ and then by HD allows the temperature to plummet 
isobarically until $\sim 100$ K, where the fragmentation mass scale of 
a few $10$ M$_{\sun}$ is imprinted.
In both the cases, temperature evolution after fragmentation
converges to the well-known track for clouds collapsing by the self-gravity 
and cooling by H$_2$ and HD (lower thin solid line; 
Uehara \& Inutsuka 2000; Nagakura \& Omukai 2005).

Next, we see the two cases starting from the filled-circle initial conditions in Fig. \ref{fig:region}, 
whose evolutionary tracks are indicated by the dash-dotted lines. 
In the higher initial temperature case ($10^4$ cm$^{-3}$, $3.6\times 10^4$ K),
the gas cools isobarically until $8000$ K as in the open-circle cases.
At $\sim 8000$ K, however, the density exceeds $\sim 10^4$ cm$^{-3}$, 
the critical density for H$_2$ to reach LTE.
For higher density, H$_2$ is rapidly dissociated collisionally 
from the excited rovibrational levels, and thus sufficient H$_2$ for cooling 
is never formed.
Consequently, the gas cannot cool below $\sim 8000$ K, and 
massive clouds with $\ga 10^5~{\rm M}_{\sun}$ are formed by fragmentation. 
Also, in the case of initial temperature somewhat lower than 8000 K
($2\times 10^5$ cm$^{-3}$, $5\times 10^3$ K), 
H$_2$ cooling is suppressed by the collisional dissociation 
and fragmentation occurs immediately producing massive clouds 
with $\ga 10^5~{\rm M}_{\sun}$, whose temperature increases to $8000$ K 
by the compressional heating in the course of gravitational collapse.
In both cases, the massive clouds thereafter collapse almost isothermally 
by the Ly$\alpha$ cooling 
until very high density ($\sim 10^{16}$ cm$^{-3}$), 
where the cloud becomes optically thick to the H$^-$ bound-free absorption 
(Omukai 2001).
Such isothermally contracting clouds do not fragment in the later phase and thus
collapse monolithically to SMSs, which eventually evolve to 
seeds of SMBHs (Bromm \& Loeb 2003; Shang, Bryan \& Haiman 2010).

As seen above, the behaviors of thermal evolution can be classified 
into two types. 
The thick solid line in Fig.~{\ref{fig:region}} corresponds to the boundary 
of initial conditions, from the above or below which subsequent 
thermal evolution bifurcates.
Namely, the post-shock conditions above the boundary 
(the hatched region) lead to the isothermal evolution at $\sim 8000$ K, 
while those below it result in the isobaric temperature decrease 
until $\la 100$ K. 
The boundary on the low-density side ($n_{\rm H,0} \la 10^4~{\rm cm}^{-3}$) 
can be fitted as 
\begin{equation}
T_0\ga 8 \times 10^3\Big( \frac{n_{\rm H, 0}}{7 \times 10^3~{\rm cm}^{-3}}\Big) ^{-1}~{\rm K}.
\label{eq:bound_low}
\end{equation}
This means that, after isobaric cooling to $8000$ K, 
if the density exceeds the H$_2$ critical value for LTE 
($\sim 10^4$ cm$^{-3}$), the gas cannot continue further isobaric compression 
and starts isothermal collapse.
For higher densities $n_{\rm H,0} \ga 10^4~{\rm cm}^{-3}$, 
the boundary is given by 
\begin{equation}
T_0\ga 5 \times 10^3\Big( \frac{n_{\rm H, 0}}{10^5~{\rm cm}^{-3}}\Big) ^{-0.1}~{\rm K}.
\label{eq:bound_high}
\end{equation}
In Section 4, we discuss physical processes 
determining the location of the boundary in more detail.

\begin{figure}
\begin{center}
\rotatebox{0}{\includegraphics[height=57mm,width=80mm]{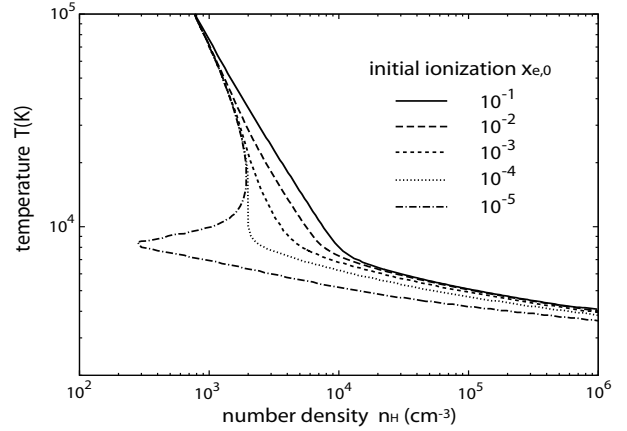}}
\end{center}
\caption{
The effect of initial ionization degree 
on the range of post-shock initial conditions 
leading to isothermal collapse by H atomic cooling 
and thus SMS formation. 
The lines present their boundaries for 
the cases with different initial ionization degrees 
of $x_{\rm e,0}=10^{-1}$ (solid), $10^{-2}$ (long-dashed), $10^{-3}$ (short-dashed),
$10^{-4}$ (dotted), and $10^{-5}$ (dash-dotted), respectively.
The molecular fraction $x_{\rm H_2,0}=10^{-6}$ 
for all cases.
}
\label{fig:e}
\end{figure}

We mention the effect of initial chemical composition on thermal evolution. 
In Fig.~\ref{fig:e}, we show the boundaries for the SMS-forming conditions 
for different initial ionization degrees ($10^{-5}\leq x_{\rm e,0}\leq 10^{-1}$).
The positions of the boundaries are almost independent of $x_{\rm e,0}$ for 
$\ga 10^4$ cm$^{-3}$, 
while the portions at $\la 10^4$ cm$^{-3}$ move to lower density 
with decreasing $x_{\rm e,0}$.
In particular, for $x_{\rm e,0}=10^{-5}$, this results in the 
spiky domain around $8000$ K extending as low as 
$\sim 3\times 10^{2}$ cm$^{-3}$.
With the higher initial ionization degree, 
the more H$_2$ is formed by the electron-catalyzed reactions 
(\ref{eq:H-}),
which results in the wider range of post-shock conditions for H$_2$ cooling, 
i.e., the smaller range for SMS formation.
The boundary for $x_{\rm e,0}\leq 10^{-2}$ asymptotically 
approaches that for higher $x_{\rm e,0}$ at $\ga 3\times 10^4$ K
since the ionization degree jumps up immediately to $\sim 10^{-1}$, 
even with smaller initial value, by effective collisional ionization 
\begin{equation}
{\rm H} + {\rm e}^-\rightarrow  {\rm H}^+ + 2~{\rm e}^-,
\label{eq:ci}
\end{equation}
in this temperature range.
We also studied the cases with different molecular fraction 
$x_{\rm H_2,0}=10^{-6}, 10^{-4}$ and $10^{-3}$, and found that 
the boundaries for SMS formation is almost independent of $x_{\rm H_2,0}$.
This is because, even with high initial value, 
H$_2$ is rapidly dissociated collisionally
for $\la 10^{4} {\rm cm^{-3}}$, and its fraction reaches 
the equilibrium one, which is independent of initial value.

\subsection{Metallicity effect}
\begin{figure}
\begin{center}
\rotatebox{0}{\includegraphics[height=60mm,width=80mm]{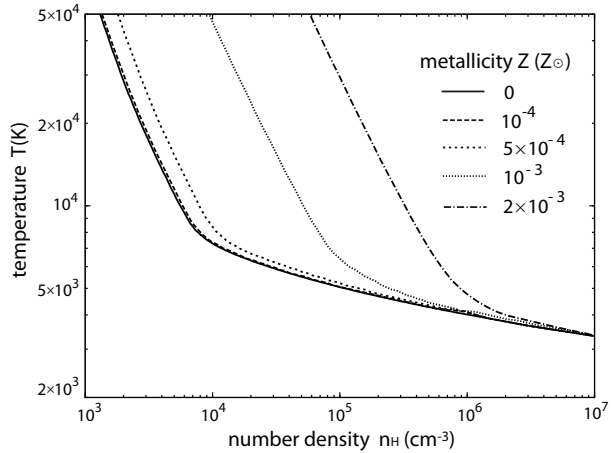}}
\end{center}
\caption{
The effect of metallicity on the range of post-shock initial conditions 
leading to SMS formation. 
The lines present their boundaries for 
the cases of metallicity $Z=0$ (solid), $10^{-4}$ (long-dashed), 
$5 \times 10^{-4}$ (short-dashed), $10^{-3}$ (dotted), $2 \times 10^{-3} 
{\rm Z}_{\sun}$ (dot-dashed), respectively.
With metallicity higher than $Z_{\rm cr}=10^{-3}~{\rm Z}_{\sun}$, 
the boundary remarkably shifts toward higher density.
The initial ionization degree and molecular fraction are 
$x_{\rm e,0}=10^{-2}$ and $x_{\rm H_2,0}=10^{-6}$, respectively.
}
\label{fig:metal}
\end{figure}

Next, we consider the cases with slight metal enrichment.
In Fig.~\ref{fig:metal}, we present the boundaries for 
SMS-forming initial conditions for various metallicities with 
$0\leq Z\leq 2\times 10^{-3}~{\rm Z}_{\sun}$.
With some metals, cooling by the 
fine-structure line emission of C$_{\rm II}$ and O$_{\rm I}$ 
can exceed that by H$_2$ and plays an important role 
in thermal evolution. 
With metallicity as low as $Z\la 5\times 10^{-4}$ Z$_{\sun}$, 
the metal cooling does not affect thermal evolution 
around 8000 K and the boundary does not move from the primordial case.
As seen in Section 3.1, from initial conditions hotter and denser than 
the boundary (i.e., the hatched region of Fig.~\ref{fig:region}), 
enough H$_2$ for cooling is not formed and only massive 
clouds are produced, which collapse isothermally thereafter.
With increasing metallicity, the metal-line cooling becomes 
able to make the gas to cool below $\sim 5000$ K isobarically even without H$_2$.
Once the temperature decreases and the collisional dissociation becomes 
ineffective, abundant H$_2$ eventually forms and the gas cools to
$\la 100$ K by its cooling.
By this additional cooling, the boundary of the SMS-forming initial conditions 
moves to higher density.
With metallicity as high as $Z\sim 10^{-3}~Z_{\sun}$, 
the cooling rate by C$_{\rm II}$ and O$_{\rm I}$
becomes comparable to the compressional heating 
at $\sim 8000$ K and $\sim 10^4$ cm$^{-3}$.
Therefore, even without help of H$_2$ cooling,
the metal cooling alone is able to lower the temperature 
to the range where the H$_2$ collisional dissociation is ineffective, 
and thus the boundary shifts to higher density.

In summary, with metallicity higher than the critical value 
$Z_{\rm cr}\sim 10^{-3}~{\rm Z}_{\sun}$, the boundary 
density becomes far higher than the typical
post-shock value by the cold accretion shock $\sim 10^3~{\rm cm}^{-3}$ 
and such an initial condition would be very difficult to be realized. 
Thus the possibility of SMS formation is strongly reduced for 
higher metallicity. 
On the other hand, as long as $Z<Z_{\rm cr}$,
the range of initial conditions for SMS formation 
remains the same as in the primordial case.

\section{Mechanism for the bifurcation of thermal evolution}
In this Section, we explain what processes are responsible for 
the bifurcation of thermal evolution in the region where 
the cold accretion shock is thermalized, and give a physical 
interpretation for the location of the bifurcation boundary
for the post-shock conditions in Figs.~\ref{fig:region} and \ref{fig:e}.

\begin{figure}
\begin{center}
\rotatebox{0}{\includegraphics[height=60mm,width=80mm]{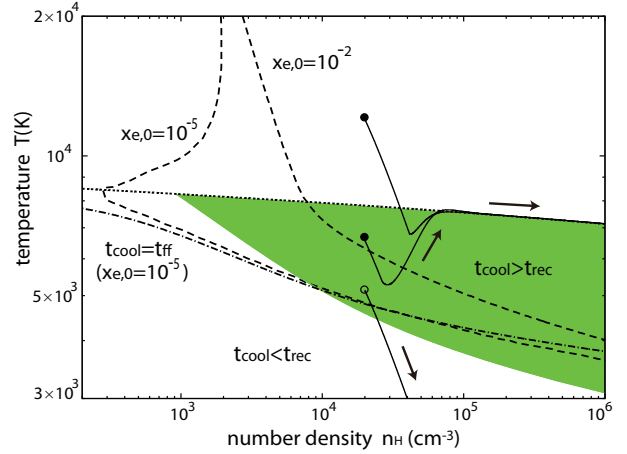}}
\end{center}
\caption{
The diagram showing relevant processes in setting the bifurcation of
the thermal evolution of the shocked gas. 
The solid lines show the evolutionary tracks of the shocked gas 
with $x_{\rm e,0}=10^{-2}$, whose post-shock conditions are 
$n_{\rm H,0}=2\times 10^4$ cm$^{-3}$ and 
$T_0=1.2\times 10^4$, $6.6\times 10^3$, and $5.1\times 10^3$ K, respectively.
Two dashed lines present the bifurcation boundaries of the SMS-forming 
initial conditions for $x_{\rm e,0}=10^{-2}$ and $10^{-5}$, respectively.
The dotted line is the same as the upper thin line in Fig.~\ref{fig:region},
which corresponds to the cut-off temperature below which the H$_2$ line-emission 
becomes the main cooling process instead of the Ly$\alpha$ emission.
In the hatched region, the ionization degree rapidly falls by effective recombination
within the cooling time of H$_2$ ($t_{\rm cool}>t_{\rm rec}$).
The dot-dashed line presents the condition of $t_{\rm ff}=t_{\rm cool}$ for $x_{\rm e}=10^{-5}$.
Here, the cooling time $t_{\rm cool}$ is evaluated by using equation (\ref{eq:H2}).
}
\label{fig:cool_rec}
\end{figure}

Efficient Ly$\alpha$ cooling drives temperature in a hot gas rapidly 
to  $\simeq 8000$ K, where its cooling rate is sharply cut off 
as the atomic hydrogen is not excited for lower temperature.
In Fig.~\ref{fig:cool_rec}, we show the cut-off temperature (dotted line),
below which H$_2$ takes over the role of the dominant coolant.
Thus, for the post-shock gas to continue 
the isobaric cooling below $8000$ K,
the H$_2$ cooling must become effective and 
keep the cooling time $t_{\rm cool}$ shorter 
than the free-fall time $t_{\rm ff}$.
The cooling time by the H$_2$ cooling is given by
\begin{equation}
t_{\rm cool}=\frac{(3/2)k_{\rm B}T}{{\cal L}_{\rm H_2}x_{\rm H_2}},
\label{eq:cool}
\end{equation}
where ${\cal L}_{\rm H_2}=\Lambda _{\rm H_2}/(n_{\rm H}x_{\rm H_2})$ 
is the cooling rate per an H$_2$ molecule (erg s$^{-1}$) and has the 
density dependence ${\cal L}_{\rm H_2}\propto (1+n_{\rm H,cr}/n_{\rm H})^{-1}$
($n_{\rm H,cr}\simeq 10^4$ cm$^{-3}$ is the H$_2$ critical density for LTE).
In conditions under consideration (below the dotted line of Fig.~\ref{fig:cool_rec}),
the H$_2$ fraction $x_{\rm H_2}$ is set by the equilibrium between 
the electron-catalyzed formation reaction and collisional dissociation reaction
\footnote[1]
{Only with high $x_{\rm e}\sim 10^{-1}$, the charge exchange reaction
(${\rm H_2}+{\rm H}^+\rightarrow {\rm H_2}^++{\rm H}$) 
becomes the main dissociation reaction of H$_2$ at $\la 10^3$ cm$^{-3}$.
However, since the bifurcation boundary for $x_{\rm e,0}=10^{-1}$ 
locates at higher density ($\ga 10^4$ cm$^{-3}$), 
the charge exchange reaction does not have any influences on the location of 
the boundary.
Thus, we adopt equation (\ref{eq:H2}) even for $x_{\rm e}\sim 10^{-1}$.}, and so
\begin{equation}
x_{\rm H_2}=\frac{k_{\rm form}}{k_{\rm cd}}x_{\rm e},
\label{eq:H2}
\end{equation}
where $k_{\rm form}$ (${\rm H}+{\rm e}^-\rightarrow {\rm H}^-+\gamma$) and
$k_{\rm cd}$ (${\rm H}_2+{\rm H}\rightarrow 3{\rm H}$)
are reaction rate coefficients for the indicated reactions, respectively.
Note that $k_{\rm cd}$ depends on the fraction of H$_2$ in excited states
and thus on the density.
The rate coefficient $k_{\rm cd}$ significantly increases with density 
near the H$_2$ critical density $n_{\rm H,cr}$,
which results in the rapid decrease of $x_{\rm H_2}$ 
by the effective collisional dissociation at $\ga 10^3$ cm$^{-3}$.

Furthermore, since $x_{\rm H_2}$ is proportional to $x_{\rm e}$,
the recombination also can be a key reaction to determine 
$t_{\rm cool}$ ($\propto x_{\rm e}^{-1}$) at $\la 8000$ K.
Since the recombination time
$t_{\rm rec}$ ($=1/\alpha _{\rm rec}n_{\rm H}x_{\rm e}$; 
$\alpha _{\rm rec}$ is the recombination rate coefficient)
has the same dependence on $x_{\rm e}$ as $t_{\rm cool}$, 
the ratio of these two timescales
becomes independent of $x_{\rm e}$ and is approximately given by
\begin{align}
\frac{t_{\rm cool}}{t_{\rm rec}}&=\frac{(3/2)n_{\rm H}k_{\rm B}T}
{{\cal L}_{\rm{H}_2}}\frac{k_{\rm cd}\alpha _{\rm rec}}{k_{\rm form}}\nonumber\\
&\simeq 0.9\left( \frac{T}{5000~{\rm K}}\right) ^{9.0}\left( \frac{n_{\rm H}}{10^4~{\rm cm}^{-3}}\right).
\label{eq:condition2}
\end{align}
The last expression above is valid for the density and temperature 
around $t_{\rm cool}\simeq t_{\rm rec}$ and $\ga 10^4$ cm$^{-3}$, 
and the large temperature dependence of equation (\ref{eq:condition2}) is due 
to that of $k_{\rm cd}$. 
In Fig.\ref{fig:cool_rec}, we show the range of the parameters satisfying 
$t_{\rm cool}>t_{\rm rec}$ (the hatched region), where the recombination 
effectively works during the isobaric contraction.
Note that the hatched region appears only at $\ga 10^3$ cm$^{-3}$,
where $k_{\rm cd}$ is enhanced significantly.
On the other hand, the ionization degree is frozen during the isobaric compression
at the density lower than the hatched region as $t_{\rm cool}<t_{\rm rec}$.
Under constant $x_{\rm e}$, the cooling time becomes shorter 
and shorter with decreasing temperature as $t_{\rm cool} \propto T^{9.7}$ (at $\simeq 5000$ K)
because the collisional dissociation is strongly suppressed for lower temperature.
Thus the efficient H$_2$ cooling and resultant isobaric evolution continue until $\sim 100$ K.

We first consider the cases with $x_{\rm e,0}\ga 10^{-4}$ and 
defer the discussion for the lower ionization cases later.
As an example, in Fig.~\ref{fig:cool_rec} we present the evolutionary tracks of the shocked gas 
with $x_{\rm e,0}=10^{-2}$ for different initial temperatures (solid lines)
and the bifurcation boundary of the SMS-forming initial condition (right dashed line). 
While the gas passes through the hatched region where $t_{\rm rec}>t_{\rm cool}$, 
the ionization degree and thus H$_2$ fraction fall rapidly. 
If such a gas runs out of H$_2$ before reaching the region where $t_{\rm cool}<t_{\rm rec}$, 
the condition for fragmentation 
$t_{\rm cool} \ga t_{\rm ff}$ is immediately satisfied and 
the clouds formed in this way collapse isothermally thereafter
(the solid lines starting from the filled circles in Fig.\ref{fig:cool_rec}).  
However, there is a small margin of the initial parameter range 
above the line $t_{\rm cool} = t_{\rm rec}$ from which the gas manages 
to maintain a small fraction of H$_2$ and can reach the region where $t_{\rm cool}<t_{\rm rec}$.
In this case, the post-shock layer can continue the isobaric contraction until $\sim 100$ K
(the solid line starting from the open circle in Fig.\ref{fig:cool_rec}).  
In summary, if the fragmentation condition $t_{\rm cool} \ga t_{\rm ff}$ is 
met in the range where $t_{\rm rec}<t_{\rm cool}$, the post-shock layer cannot 
cool further and clouds formed at this moment begin the isothermal collapse.
The set of initial conditions from which the evolutionary tracks meet the condition 
$t_{\rm ff}=t_{\rm cool}$ just on the line $t_{\rm rec}=t_{\rm cool}$ 
corresponds to the boundary for the SMS formation in 
Fig.~\ref{fig:region} in the high density regime 
(i.e., equation \ref{eq:bound_high}).
Note that the boundary given by equation (\ref{eq:bound_high}) reflects 
the density-temperature relation of $t_{\rm rec}=t_{\rm cool}$, 
which is mainly determined by the temperature dependence of $k_{\rm cd}$.
Due to the strong dependence of $k_{\rm cd}$ on temperature,
the H$_2$ is collisionally dissociated very efficiently for 
the temperature higher than given by equation (\ref{eq:bound_high}).
On the other hand, if the density is lower than the H$_2$ critical density
$n_{\rm H,cr}$ ($\sim 10^4$ cm$^{-3}$) after cooling isobarically to $8000$ K, 
the gas continues further isobaric contraction to  several $100$ K 
by the ineffective collisional dissociation of H$_2$.
This explains the fact that the low-density side of the boundary 
of SMS-forming parameters in Fig.~\ref{fig:region} (i.e., equation \ref{eq:bound_low})
corresponds to the isobaric contraction track whose density at $\simeq 8000$ K is 
$\sim 10^4$ cm$^{-3}$.

Next, we consider the low ionization cases with $x_{\rm e,0}\la10^{-4}$.
As seen in Fig.~\ref{fig:cool_rec}, the portion of the bifurcation boundary (left solid line) 
for $x_{\rm e,0}=10^{-5}$ locates in the region $t_{\rm rec}>t_{\rm cool}$, 
where $x_{\rm e}$ is frozen during the isobaric compression.
Therefore, if the cooling condition $t_{\rm cool}<t_{\rm ff}$ is initially satisfied 
in the post-shock layer, the gas continues to cool isobarically until $\la 100$ K,
where it produces $\sim 10$ M$_{\sun}$ fragments.
Thus, the boundary of the SMS-forming condition is simply 
given by the requirement $t_{\rm ff} \la t_{\rm cool}$ for 
their initial values without the need for considering the recombination effect. 
The dot-dashed line in Fig.~\ref{fig:cool_rec} presents the condition 
$t_{\rm ff}=t_{\rm cool}$ for $x_{\rm e}=10^{-5}$ and in fact coincides with 
the boundary on the low-temperature side ($\la 8000$ K).

\section{Conclusion and Discussion}
In this paper, we have proposed a new scenario for supermassive star (SMS) formation 
in central hot and dense regions of the halos formed by the cold accretion shocks 
in the first galaxy formation.
Since the gas cools effectively by Ly$\alpha $ emission in halos with 
virial temperature $T_{\rm vir}\ga 10^4$ K, 
location of the accretion shock does not stay at the virial radius 
and shrinks inward. 
The gas instead flows supersonically along cold and dense filaments 
to the central region of the first galaxy, 
where the flows collide each other to produce a hot ($\ga 10^4$ K) and dense 
($\ga 10^3$ cm$^{-3}$) material by a shock 
(Birnboim \& Dekel 2003; Kere{\v s} et al. 2005; Dekel \& Birnboim 2006; 
Wise \& Abel 2007; Greif et al. 2008; Wise et al. 2008; Dekel et al. 2009; 
Bromm \& Yoshida 2011).
We have calculated thermal evolution in such a hot and dense region formed by 
the cold accretion shock.
For $\ga 8000$ K, the efficient Ly$\alpha$ cooling allows 
the post-shock gas to cool and to contract isobarically 
at the value of ram pressure from the shock front.
To continue the isobaric cooling also below $8000$ K, 
abundant H$_2$ needs to be formed and its cooling must be effective.
If the density at $\simeq 8000$ K is high enough ($\ga 10^4$ cm$^{-3}$) 
to make the H$_2$ rovibrational levels to reach the local thermodynamic equilibrium,
the H$_2$ is dissociated effectively by the collisional reaction 
from excited levels, which suppresses the cooling to lower temperature by H$_2$.
At this epoch, gravitational instability of the post-shock layer 
produces massive fragments with $\ga 10^5$ M$_{\sun}$, 
which subsequently collapse isothermally at $\sim 8000$ K by the Ly$\alpha $ cooling.
We have studied thermal evolution of the post-shock gas for a wide range 
of initial conditions
($10^2$ cm$^{-3}<n_{\rm H,0}<10^7$ cm$^{-3}$ and $3000$ K$<T_0<10^5$ K) 
and have pinned down the conditions leading to the isothermal collapse 
(the hatched region in Fig.~\ref{fig:region}):
$T_0\ga 6000~(n_{\rm H,0}/10^4~{\rm cm}^{-3})^{-1}$ K 
for $n_{\rm H,0} \la 10^4~{\rm cm}^{-3}$ and 
$T_0\ga 5000-6000$ K for $n_{\rm H,0} \ga 10^4~{\rm cm}^{-3}$,
for the pre-shock ionization degree at $x_{\rm e,0}=10^{-2}$.
Since H$_2$ is formed by the electron-catalyzed reactions
(eq. \ref{eq:H-}),
the above condition depends somewhat on the initial ionization degree (see Fig.~\ref{fig:e}): 
for smaller $x_{\rm e,0}$, the domain of initial conditions leading to the isothermal 
collapse extends towards lower density. 
Those massive clouds continue isothermal collapse 
until very high density $\sim 10^{16}$ cm$^{-3}$, where they become optically 
thick to the H$^-$ bound-free absorption (Omukai 2001).
The clouds are supposed to collapse directly to SMSs without further 
fragmentation (e.g., Bromm \& Loeb 2003; Regan \& Haehnelt 2009a, b; 
Shang, Bryan, \& Haiman 2010).
Eventually, the SMSs collapse by the post-Newtonian instability, 
swallowing most of their material, to become 
seeds of SMBHs (Shibata \& Shapiro 2002).

The first galaxies may be enriched with metals to some extent as well as  
dusts dispersed by supernova (SN) explosions of previous generations of stars.
With high enough metallicity, the gas can cool to low temperature 
($\sim 100$ K) by metal-line cooling alone even without H$_2$.
In this case, the SMS formation would be strongly suppressed. 
We have then repeated the same analysis by considering as well as
the metal cooling by C$_{\rm II}$ and O$_{\rm I}$. 
We have found that as long as the metallicity is lower than 
$Z_{\rm cr} \simeq 10^{-3}$ Z$_{\sun}$, 
the metal-line cooling does not change the condition for SMS formation 
from that in the primordial case (see Fig.~\ref{fig:metal}).
According to some cosmological simulations of the assembly of first galaxies
(Greif et al. 2010; Wise et al. 2012),
the dense gas at the center of galaxies is uniformly enriched to 
$\sim 10^{-3}$ Z$_{\sun}$ by a pair instability SN of 
massive population III stars with 
$140~{\rm M}_{\sun}\la {\rm M}\la 260~{\rm M}_{\sun}$.
On the other hand, typical population III stars are recently considered 
to be less massive $\sim 40~{\rm M}_{\sun}$ 
(Hosokawa et al. 2011; Stacy, Greif \& Bromm 2011)
and end their lives as ordinary core collapse SNe.
In this case, the resultant metallicity reduces by a factor of $\sim $10 
(Heger \& Woosley 2002; Nomoto et al. 2006)
and thus becomes lower than the critical metallicity for SMS formation 
we estimated.

We here stress that our scenario naturally explains 
the formation of seed BHs in the conditions of first-galaxy formation 
without invoking extremely strong UV radiation as envisaged 
in the previous scenario. 
The necessary condition for SMS formation in halos with 
$T_{\rm vir}\ga 10^4$ K is the isothermal collapse by the atomic cooling 
as a result of suppression of the H$_2$ cooling in the entire density range.   
So far, as a mechanism to suppress H$_2$ cooling, photodissociation by FUV 
radiation has been considered.
In this scenario, however, extremely strong FUV intensity
$J_{21}^{\rm LW}\ga 10^2-10^3$ (in unit of $10^{-21}$ erg s$^{-1}$ cm$^{-2}$ 
Hz$^{-1}$ sr$^{-1}$) is required to quench the H$_2$ cooling 
(Omukai 2001; Bromm \& Loeb 2003; Shang, Bryan \& Haiman 2010), and 
halos irradiated by such intense radiation are extremely rare
($\la 10^{-6}$ at $z\sim 10$; Dijkstra et al. 2008).
Moreover, if external ionization by cosmic rays or X-rays, 
which promotes the H$_2$ formation, is present as well, 
the FUV intensity needed for SMBH formation is elevated. 
There would be little possibility ($\ll 10^{-6}$) for such 
an intense FUV field to be realized in any haloes and thus 
the SMS formation could strongly suppressed (Inayoshi \& Omukai 2011).
On the other hand, in our scenario, collisional dissociation, rather than 
photodissociation, suppresses the H$_2$ cooling.
Thus, even without FUV radiation, which has been considered 
to be indispensable previously, 
the isothermal collapse and thus SMS formation can be realized 
as long as the right condition is met for the cold accretion shock. 
Note, however, this mechanism for SMS formation cannot 
operate in all first galaxies since SMBHs are rare objects.
If we use the number density of halos with mass $\sim 10^8~{\rm M}_{\sun}$ 
at $z\sim 10$, $\sim 10~{\rm Mpc}^{-3}$ (comoving)
and assume each of them had a SMS of $\sim 10^5~{\rm M}_{\sun}$, 
the predicted mass density 
of SMBHs $\sim 10^6~{\rm M}_{\sun}~{\rm Mpc}^{-3}$ (comoving)
will exceed the total present-day BH mass density 
$\sim 3\times 10^5~{\rm M}_{\sun}~{\rm Mpc}^{-3}$ estimated by
Yu \& Tremaine (2002).
Therefore, our conditions for SMS formation 
would be satisfied only in a small fraction of first galaxies 
or some other processes, e.g., turbulent fragmentation, lack of accreting material etc., 
suppress this mechanism to work for avoiding overproduction of BHs.

We here briefly discuss the effect of dust cooling, 
which has not been considered in this paper. 
If the depletion factor of metals to dust grains is as high as 
the present-day Galactic value $f_{\rm dep}\simeq 0.5$, 
the thermal evolution deviates from the isothermal one 
at $\ga10^{10}$ cm$^{-3}$ due to the dust cooling,
if metallicity is higher than $Z_{\rm cr, dust}\simeq 10^{-5}~{\rm Z}_{\sun}$
(Omukai, Schneider \& Haiman 2008).
Although this critical metallicity $Z_{\rm cr,dust}$ is 
smaller than the critical value due to metal-line cooling 
$Z_{\rm cr}\simeq 10^{-3}~{\rm Z}_{\sun}$ by two orders of magnitude,  
the depletion factor in the first-galaxy forming environment 
is highly uncertain.
According to theoretical models of dust formation and destruction 
in the first SNe, typically only a few \% of dust formed at the explosion survives, 
after being swept by the reverse shock,
depending on the ambient density 
(Nozawa, Kozasa \& Habe 2006; Bianchi \& Schneider 2007).
For example, with $f_{\rm dep}\simeq 0.05$, 
the critical metallicity becomes 
$Z_{\rm cr,dust} \simeq 10^{-4}~{\rm Z}_{\sun}$, 
which makes the constraint on metal pollution less severe.
In any case, to predict whether the isothermal collapse continues
in spite of metal enrichment,
we need more accurate knowledge of the depletion factor $f_{\rm dep}$. 

In this paper, we consider the hot and dense central regions 
owing to shocks by the cold accretion flows in the forming first galaxies.  
Likewise, a galaxy merging event drives inflows, creating 
a similar environment around the galaxy center (e.g., Mayer et al. 2010). 
If the shocked region satisfies our post-shock criterion for 
the H$_2$ collisional dissociation, the SMS formation is expected 
also in this case. 
In merging galaxies, however, star formation and also metal 
enrichment are expected to 
have already proceeded significantly.
Therefore, in the case of the inflows by galaxy merger, SMS formation is 
probably prohibited by the metal-cooling effect. 
The cold accretion shocks in the first galaxy formation would
provide more easily the suitable conditions for SMS formation.
Recently, assuming seed BHs with $\sim 10^5~{\rm M}_{\sun}$, 
Di Matteo et al. (2011) and Khandai et al. (2011) discussed 
their growth by the cold accretion flows 
in the process of the first galaxy formation.
Their results demonstrate that the cold flows are less susceptible 
to feedbacks from the growing BHs and high accretion rate 
is maintained until mass of the galaxy reaches $\ga 10^{12}~{\rm M}_{\sun}$, 
where the cold mode of accretion turns the usual hot virialization mode. 
As a result, the BH is able to grow to $\ga 10^9~{\rm M}_{\sun}$ by $z\sim 6$.
Our scenario for the SMS formation provides a mechanism for seeding BHs of 
$\sim 10^5~{\rm M}_{\sun}$ in forming galaxies, which has been assumed in studies of 
Di Matteo et al. (2011) and Khandai et al. (2011), while their results complementary 
demonstrated those seed BHs can in fact grow to SMBHs. 

Finally, we remark remaining issues to be explored.
In our scenario, the physical condition in the post-shock gas 
(especially, its density) is crucial for the SMS formation.
We have considered a range of density ($10^2-10^7 {\rm cm}^{-3}$) 
and temperature ($3000-10^5$ K) as the post-shock conditions.
Currently, we only know the typical values of those parameters 
(Greif et al. 2008; Wise et al. 2008), but are still lacking knowledge 
of the relationship between the post-shock conditions and formation conditions 
(i.e., mass, virialization epoch, etc.) of first galaxies. 
In addition, our assumption that the enough mass supply for Jeans instability 
is available through the streaming flow is need further investigation. 
Safranek-Shrader et al. (2010) evaluated the amount of accreted gas to be 
$\sim 10^{5}M_{\sun}$, which is inhomogeneously organized by turbulence 
(e.g., Wise \& Abel 2007; Greif et al. 2008).
Therefore, the outcome of the shocked material in most of halos 
could be numerous small sub-regions, rather than a massive 
layer envisaged in this paper.  
On the other hand, even if the turbulent motion dominates, 
a gravitational unstable cloud with $\sim 10^5$ M$_{\sun}$ could form 
in the center and collapse subsequently (Wise et al. 2008). 
As a future project, we need to study in what halos the SMS forming 
conditions are satisfied by way of realistic cosmological simulations, 
including e.g., molecular cooling and the radiative and chemical feedbacks, 
for evaluating more quantitatively the feasibility of SMS formation in the first galaxies.

In this paper, we have supposed that, 
for the SMS formation, massive clouds must collapse isothermally without fragmentation.
In fact, three-dimensional hydrodynamical simulations by Bromm \& Loeb (2003) 
confirmed in some cases that the cloud collapsing isothermally by the atomic cooling
does not experience fragmentation at least in the range $\la 10^9$ cm$^{-3}$ 
and, consequently, a supermassive clump forms at the center. 
Even with some angular momentum, fragmentation resulted 
at most in a binary system in their calculation. 
However, depending on such initial conditions as degrees of 
rotation or turbulence, the clouds would fragment into less massive clumps 
during the isothermal collapse. 
As a future study, it is awaited to clarify
the conditions under which the clouds elude fragmentation 
by way of three-dimensional hydrodynamical calculation.  

\section*{Acknowledgments}
We would like to thank Takashi Nakamura for his continuous 
encouragement, and Takashi Hosokawa and Sanemichi Takahashi 
for the fruitful discussion.  
We also thank an anonymous referee for a careful reading of 
the manuscript and for constructive criticism that improved this paper.
This work is in part supported by the Grants-in-Aid by the Ministry 
of Education, Culture, and Science of Japan (23$\cdot $838 KI; 2168407 
and 21244021 KO).

\bsp

\label{lastpage}

\end{document}